\documentstyle[12pt,aaspp4,flushrt]{article}
  
\def\etal{{\it et $\!$al.\,}}

\begin{document}
 
\title{Coherently-Dedispersed Polarimetry of Millisecond Pulsars}
\author{I. H. Stairs\altaffilmark{1,2}, S. E. Thorsett\altaffilmark{2}, 
and F. Camilo\altaffilmark{1}}
\altaffiltext{1}{The University of Manchester, Nuffield Radio Astronomy
Laboratories, Jodrell Bank, Macclesfield, Cheshire SK11 9DL UK;
is@jb.\-man.\-ac.\-uk, fc@jb.\-man.\-ac.\-uk}
\altaffiltext{2}{Joseph Henry Laboratories and Physics Department,
       Princeton University, Princeton, NJ 08544; 
steve@\-pulsar.\-princeton.\-edu}
\vspace*{0.1in}
\centerline{{\it The Astrophysical Journal Supplement Series} {\bf 123}, 627 (1999)}
 
\begin{abstract}

We present a large sample of high-precision, coherently-dedispersed
polarization profiles of millisecond pulsars (MSPs) at frequencies between
410 and 1414\,MHz.  These data include the first polarimetric
observations of several of the pulsars, and the first low-frequency
polarization profiles for others.  Our observations support previous
suggestions that the pulse shapes and polarimetry of MSPs are more
complex than those of their slower relatives.  An immediate conclusion
is that polarimetry-based classification schemes proposed for young
pulsars are of only limited use when applied to millisecond pulsars.
\end{abstract}
 
 
\section{Introduction}\label{sec:intro}

The polarization characteristics of a pulsar yield clues to its
magnetic geometry and emission mechanism.  Many pulse profiles are
highly linearly polarized and have a clear position-angle swing across
part or all of the period.  For slow pulsars, this position-angle
swing often follows the characteristic S-shaped curve of the
``rotating vector model'' (RVM) introduced by Radhakrishnan and Cooke
(1969). \nocite{rc69a} In this description, the magnetic dipole axis
is inclined at an angle $\alpha$ from the spin axis, and makes angle
$\beta$ with the line of sight from Earth.  The angle of between the
line of sight and the pulsar spin axis is $\zeta=\alpha+\beta$.  The
position angle $\psi$ can then be written as a function of pulse phase
$\phi$:
\begin{equation}\label{eqn:rvm}
\tan(\psi(\phi) - \psi_{0})\, =\, \frac{\sin \alpha \sin(\phi -
\phi_0)}
{\cos\alpha\sin\zeta - \sin\alpha\cos\zeta\cos(\phi -
\phi_0)}
\end{equation} 
where $\psi_0$ is a constant offset and $\phi_0$ is the phase of
steepest position-angle swing, given by:
\begin{equation}\label{eqn:rvmmax}
\left(\frac{d\psi}{d\phi}\right)_{\phi_0}\, =\,
\frac{\sin\alpha}{\sin\beta}. 
\end{equation}

The rotating vector model, along with the observed linear and circular
polarization properties of many pulse profiles, has led to the
formulation of successful phenomenological classification schemes for
slow pulsars based on pulse morphology and polarimetry (e.g.,
\cite{ran83,lm88}). 

To date, limited polarimetric studies of the faster millisecond
pulsars (MSPs) have suggested that their pulse components do not fit
easily into these classification schemes, and frequently have more
complex polarimetric profiles than slow pulsars
(\cite{ts90,nms+97,xkj+98}).

In this paper we present polarization profiles for several
millisecond pulsars.  These profiles represent the first polarimetric
observations of many of these pulsars, and for others, the first
low-frequency polarization profiles.  There is considerable overlap
between the present set of pulsars and those studied by Xilouris \etal
(1998) at 1410\,MHz using the 100\,m telescope at
Effelsberg, Germany; our study provides a needed complement to these
higher-frequency observations.  Recently, Sallmen (1998)
\nocite{sal98} has studied several of the pulsars presented here,
conducting observations at the National Radio Astronomy Observatory,
Green Bank, W.V. and Effelsberg, Germany. Where applicable, we compare
our results to these studies.

\section{Observations}\label{sec:obs}

Observations were made using the 76-m telescope at the Nuffield Radio
Astronomy Laboratories, Jodrell Bank, U.K., between 1997 January and
1997 July.  Data were obtained at 410\,MHz, 610\,MHz, and 1414\,MHz,
with observing bandwidths of 5\,MHz at 410\,MHz and 610\,MHz, and
10\,MHz at 1414\,MHz.  All data were acquired with the Princeton
Mark~IV observing system.  This instrument implements the coherent
dispersion removal technique described by Hankins and Rickett
(1975). \nocite{hr75} A fast analog-to-digital converter samples
quadrature components of the received telescope voltages in the two
orthogonal circular polarization channels, $L$ and $R$, employing
either 2-bit sampling across 10\,MHz of bandwidth, or 4-bit sampling
across 5\,MHz.  The data are stored on magnetic tape for off-line
processing.  Coherent dedispersion is performed in software, using a
special-purpose parallel processor.  The data stream is typically
split into either two or four sub-bands during processing.  The self-
and cross-products $|L|^2$, $|R|^2$, Re($L^*R$) and Im($L^*R$) are
formed from the dedispersed time-series in each sub-band, then folded
modulo the predicted topocentric pulse period.  The full set of Stokes
parameters are readily formed from these products.  A complete
description of the instrument, along with preliminary polarimetric
results, has been published elsewhere (\cite{sta98}).

Fluxes were calibrated using an injected noise signal, the strength of
which was determined by comparison with a bright continuum source of
known flux.  The reference sources used varied between observing
sessions, but were drawn from the following list: 3C48, with an
assumed flux of 15.8\,Jy at 1414\,MHz; 3C123, with an assumed flux of
117.6\,Jy at 410\,MHz, 91.8\,Jy at 610\,MHz, and 48.4\,Jy at
1414\,MHz; 3C147, with an assumed flux of 22.3\,Jy at 1414\,MHz; and
3C353, with an assumed flux of 129.4\,Jy at 410\,MHz.  The
uncertainties in the calibration are approximately 10\%.  The noise
calibrator was switched on and off at (typically) 3-minute intervals
throughout each 30-minute observation.

The flux in the $|R|^2$ channel was determined using the differences
in the off-pulse baseline and the strength of the noise calibrator.
The spectrum of the $|L|^2$ channel was found to be more variable
across the observing bandwidth, making a similar calculation less
reliable for this channel.  Instead, the $|L|^2$ calibration factor
was calculated so that the ratio of the off-pulse baselines in $|L|^2$
and $|R|^2$ was equal to the ratio of the known total system
temperatures, i.e., the receiver temperatures corrected for telescope
elevation angle and sky temperature in the direction of the pulsar
($T_{\rm sys} = T_{\rm rec} + T_{\rm sky} + T_{\rm
ground\,spillover}$). The square root of the product of the $|L|^2$
and $|R|^2$ calibration factors was used to calibrate the
cross-product terms.

The four Stokes parameters, $S$, $Q$, $U$, and $V$, were obtained from
the calibrated products.  A small DC bias was removed from the
cross-terms, $Q$ and $U$; these parameters were then corrected for the
parallactic rotation of the feed during tracking using the procedure
outlined by Rankin \etal (1975). \nocite{rcs75} The corrected terms,
$Q^{\prime}$ and $U^{\prime}$, yielded the linearly
polarized power $L_{\rm p} = \sqrt{Q^{\prime\,2} + U^{\prime\,2}}$ and
the source position angle $\psi = - \frac{1}{2}
\tan^{-1}(U^{\prime}/Q^{\prime})$.  We followed the recommendation of
Damour and Taylor (1992) \nocite{dt92} in taking the angle $\psi$ to
increase clockwise on the sky.  A position-angle offset was fitted out
between sub-bands and between profiles taken on different days.

\section{Observations of PSR~B1929$+$10}\label{sec:1929}

\begin{figure}
    \plotone{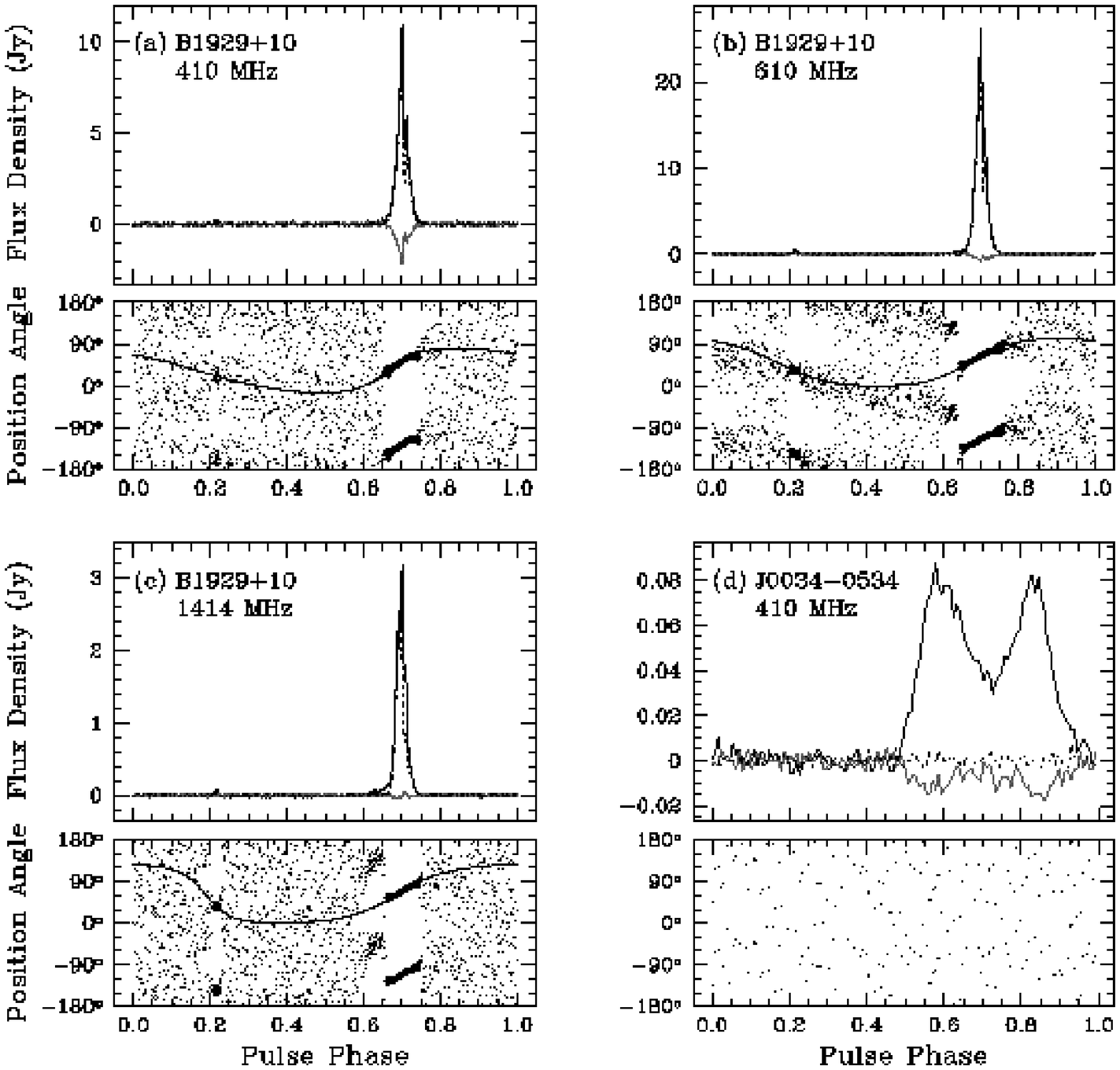}
    \figcaption{Pulse profiles for (a) PSR B1929+10 at 410\,MHz,
	(b) PSR B1929+10 at 610\,MHz,
	(c) PSR B1929+10 at 1414\,MHz,
	(d) PSR J0034$-$0534 at 410\,MHz.
	The solid black line represents the total intensity S, the
dotted black line the linear intensity L, and the solid gray line the
circular intensity V.  The position angle $\psi$ is plotted twice for
clarity; larger points are used when L is more significant.  The
curves indicate the fits to the rotating vector model.
    \label{fig:f1929}}
\end{figure}

In order to monitor the quality of our polarimetry and flux
calibration, we frequently observed the strong, slow pulsar
B1929$+$10.  This pulsar has been previously studied by several
different observers (\cite{lm88,phi90,rr97}); it is known to have high
linear polarization and very little circular intensity.  An error in
our calibration procedure would therefore be immediately apparent in
its polarization profile.  Our profiles at 410\,MHz, 610\,MHz, and
1414\,MHz are presented in Figures~\ref{fig:f1929}(a)-\ref{fig:f1929}(c); they
agree well with those reported in the literature.  We therefore
consider our calibration procedure correct to within the estimated
10\% uncertainty in the injected noise levels.  We note, however, that
at 410\,MHz the circular intensity profile appears similar to the
total power in shape, suggesting a small miscalibration in $|L|^2$ and
$|R|^2$.  The parameters derived from our RVM fits to the position
angle swing of this pulsar are given in Table~\ref{tab:1929}.

In principle, observations of B1929$+$10 can be used to correct for
any ellipticity or non-orthogonality in the nominally circular antenna
feeds (\cite{sti82,mck92,xil91}).  Such imperfections result in mixing
of, respectively, linear power into circular and total intensity into
linear, with the coupling varying sinusoidally as a function of the
incident position angle.  Gould (1994) \nocite{gou94} finds that
non-orthogonality in the 610\,MHz feeds at Jodrell Bank affects the
linear intensity at the level of a few percent; an effect of this
magnitude would not be readily separable from calibration effects in
our data.

In our 610\,MHz and 1414\,MHz profiles for B1929$+$10, the position
angle exhibits a $90^\circ$ shift near the leading edge of the main
pulse.  Such shifts are common to many pulsar profiles.  They are
accompanied by a null in the linear polarization intensity, and can be
explained by a switch between two competing orthogonal emission modes
(\cite{scr+84}).  These orthogonal mode changes can be removed before
fitting an RVM model to the position angle data.

\section{Polarimetry of Millisecond Pulsars}\label{sec:disc}

We now present the profiles of the millisecond pulsars observed at
Jodrell Bank, and discuss the more important measurements.
Table~\ref{tab:ppdot} gives the period, period derivative, magnetic
field and characteristic age for each pulsar observed, and
Table~\ref{tab:res} lists the time resolutions and integration times
for each of the displayed profiles.  Table~\ref{tab:pol} lists the
$10\%$ and $50\%$ widths, the effective pulse width, defined as the
area under the pulse profile divided by the peak height, along with
the average magnitudes of the linear and circular intensities and mean
fluxes for each of the pulsars in our study.  The $10\%$ and $50\%$
pulse widths are obtained by linear interpolation of the profiles,
yielding errors the size of a few phase bins.  The calibration
uncertainties lead to errors of approximately $10\%$ in the average
linear and circular intensities and fluxes.  For those pulsars for
which a rotating vector model fit was possible, the results are given
in Table~\ref{tab:rvm}.

\subsection{PSR~J0218$+$4232}\label{sec:0218}

PSR~J0218$+$4232 was initially discovered in a survey for
steep-spectrum point sources; the pulsed emission is on
average only one-half the flux observed from the point source
(\cite{nbf+95}).  A significant level of unpulsed emission is
extremely unusual in radio pulsars.  When considered with the fact that
there are pulsed components across the profile, this emission suggests
that the pulsar is an aligned rotator with a broad beam.

Our RVM fits support the classification as a nearly aligned rotator,
with magnetic inclinations consistent with $0^\circ$ at both 410 and
610\,MHz.  It should be noted that if the continuum emission is
strongly polarized, this pedestal could affect our measurement of the
position angle of the pulsed component.  The impact parameters from
our RVM fits have very large uncertainties, but are consistent with a
line-of-sight inclination of $90^\circ$, possible if the pulsar beam
is extremely broad.  This binary system is therefore a candidate for
the measurement of Shapiro delay, an effect seen in only one
neutron-star--white-dwarf system to date (\cite{rt91a}).

\subsection{PSR~J0613$-$0200}\label{sec:0613}

At both observing frequencies a large fraction of circularly polarized
power is evident in the main pulse but not in the two precursors.
There is significant linear polarization in the main pulse, but the
position angle does not follow a sweep consistent with the RVM.  Our
profiles are consistent with those reported by Backer at 780\,MHz
(\cite{bac95b}) and Sallmen at 575\,MHz (\cite{sal98}).  Observations
at 21\,cm at Bonn (\cite{xkj+98}) and Green Bank (\cite{sal98}) find a
very different and nearly completely unpolarized pulse profile: what
we label as the main pulse becomes a trailing pulse, while the second,
larger, precursor becomes dominant.  Thus the high-frequency profile
is similar to a triple pulse.  The lack of profile development between
410 and 780\,MHz and the completely different morphology at 1400\,MHz
imply a frequency development very different from that predicted by
the empirical model.

\subsection{PSR~J0621$+$1002}\label{sec:0621}

This pulsar has a white-dwarf companion of minimum mass $0.45~M_\odot$,
higher than that of most white-dwarf pulsar companions, but still far
below the mass of a neutron star.  As helium flash occurs at stellar
core masses above $0.45~M_\odot$ (e.g., Kippenhahn and Weigert 1990),
\nocite{kw90a} the companion is likely to be a carbon-oxygen white
dwarf.  The system is classified as being one of the few
``intermediate-mass binary pulsars (IMBPs)''; its evolutionary history
is thought to have been different from that of lower-mass systems.  In 
particular, the mass accretion history may have been different from
that seen in the low-mass systems, possibly leading to difference in
the observed emission geometry.

This pulsar exhibits very low levels of polarization, with a
relatively flat position-angle swing.  Camilo \etal (1996) report that
the pulse shape does not change between 370\,MHz and 1.4\,GHz, and our
results support this finding.  For slow pulsars it is common for the
spacing between pulse components to decrease with increasing
frequency; this pulsar clearly does not follow the same frequency
evolution.

\subsection{PSR~J1012$+$5307}\label{sec:1012}

This pulsar has a white dwarf companion for which the radial velocity
is measured; the mass ratio resulting from the observations is
$m_1/m_2 = 9.5\pm0.5$ (\cite{vker98}).  Assuming a neutron-star mass
of $1.35~M_\odot$, the timing mass function yields an expected orbital
inclination angle of $\approx 50^\circ$.  From evolutionary arguments
(e.g., Bhattacharya and van den Heuvel 1991) \nocite{bv91}, the spin
axis is likely to be nearly aligned with the orbital angular momentum
axis.  The large width of the pulse profile and the overall similarity
of the position angle swing to that of PSR~J0218+4232 lead us to
expect,  also in this case, a nearly-aligned rotator with a wide beam
and large impact parameter.

The 610-MHz profile exhibits multiple components across the period, as
well as strong linear and moderate circular polarization.  Similar
results are found at 575\,MHz by Sallmen (1998). \nocite{sal98}
Xilouris \etal (1998) point out that equal position angle slopes,
though with opposite sign, for the main pulse and interpulse would
argue for the classification of the pulsar as an orthogonal rotator.
Unfortunately, while there is a small slope in the main pulse
position-angle swing, the swing in the interpulse region is nearly
flat, making this test impractical.  Our RVM fits to the shallow
position-angle swing indicate that the magnetic inclination is instead
very close to $0^\circ$, supporting the classification as an aligned
rotator.  The impact parameter is not well-determined.

\subsection{PSR~J1022$+$1001}\label{sec:1022}

PSR~J1022$+$1001 is another IMBP, with a companion mass of
approximately $0.87~M_\odot$ \linebreak (\cite{cnst96}).  The profile is
narrow and highly polarized at all frequencies.  It is difficult to
comment on the frequency evolution of the profile, as it is known to
exhibit variations, on the time-scale of tens of minutes, in which the
leading component becomes larger or smaller than the apparently more
stable trailing component (\cite{kxc+99}).

The strongest linear polarization is associated with the trailing
component.  Despite the roughly S-shaped position-angle swing, the RVM
model is not a very good fit except at 1414\,MHz, where we find an
impact parameter of $|\beta| = 7.1\pm0.4^\circ$, consistent with the
result reported in Xilouris \etal (1998).  The narrow pulse makes it
difficult to constrain the magnetic inclination and hence the
line-of-sight angle $\zeta$.  It appears that both $\alpha$ and
$\zeta$ are approximately $90^\circ$; the direction of the off-pulse RVM
swing is determined by whether $\zeta$ falls above or below $90^\circ$.

At the two lower observing frequencies, there appears to be a
transition in the shape of the position-angle swing, associated with a
dip between two peaks of the linear polarization.  This ``notch''
makes it difficult to fit a single RVM model to the data.  At
610\,MHz, we obtain a possible fit with impact parameter $|\beta| =
4.9\pm1.8^\circ$, within error of the value obtained at 1414\,MHz.

\subsection{PSR~J1713$+$0747}\label{sec:1713}

The profiles for this pulsar show considerable frequency development.
The wider pulse shape at 410\,MHz is a change in morphology; it is not
consistent with broadening due to scattering by the interstellar
medium.  There is some circular polarization at all frequencies,
indicative of a central core component.  The strong linear intensity
is accompanied by a shallow position angle curve with multiple
orthogonal mode changes.  Similar results at 575\,MHz and 1410\,MHz
are found by Sallmen (1998).

\subsection{PSR~J1730$-$2304}\label{sec:1730}

At our observing frequency of 610\,MHz, the linear polarization is
rather weak.  Our data were collected on three different days; we find
some evidence for mode-changing in the total intensity as reported by
Xilouris \etal (1998) and Kramer \etal (1998). \nocite{kxl+98}  The
sense-reversing circular intensity also agrees with these results,
supporting the hypothesis of a central core profile component.

\subsection{PSR~B1744$-$24A}\label{sec:ter5}

\begin{figure}
    \plotone{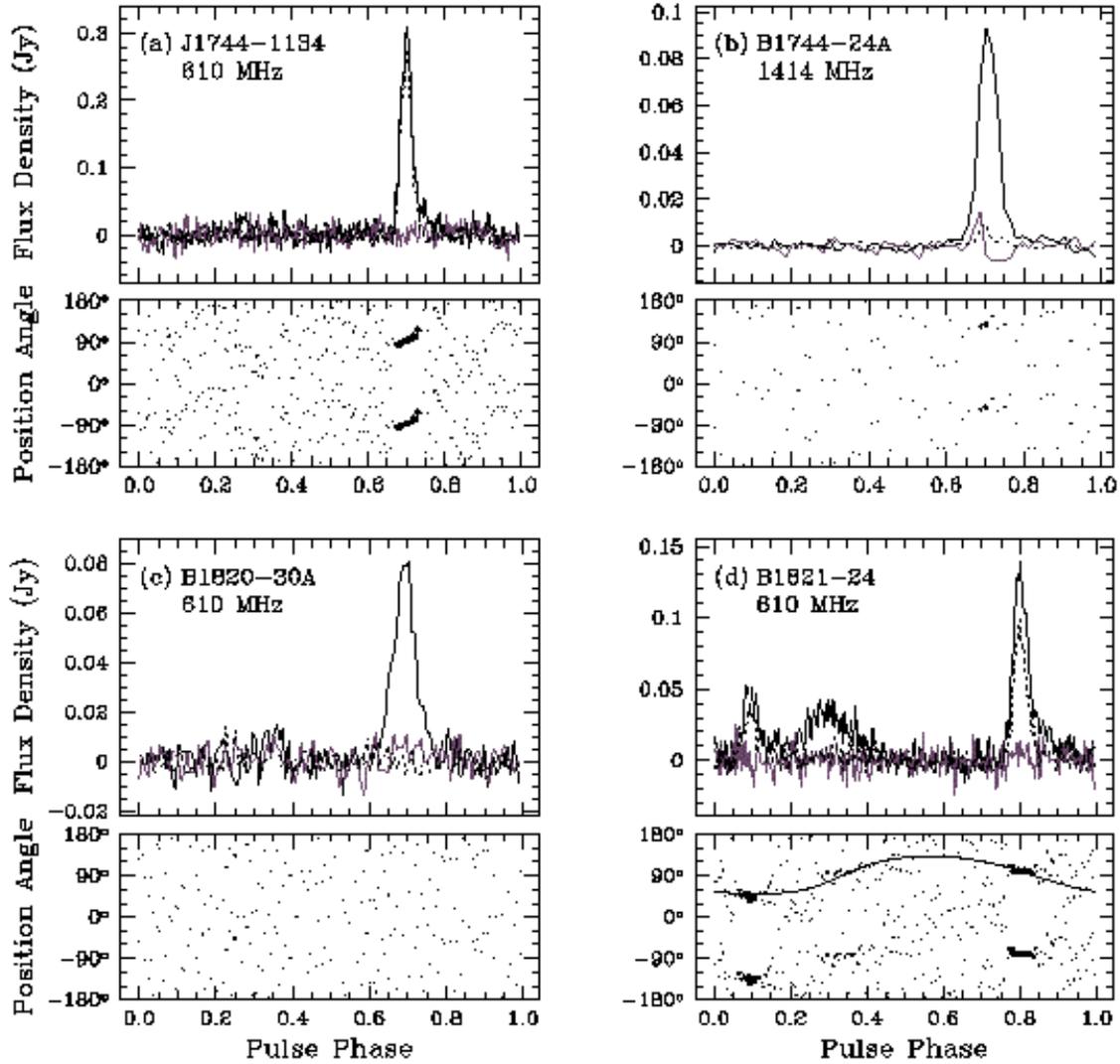}
    \figcaption
    {Pulse profiles for (a) PSR J1744$-$1134 at 610\,MHz,
	(b) PSR B1744$-$24A at 1414\,MHz,
	(c) PSR B1820$-$30A at 610\,MHz,
	(d) PSR B1821$-$24 at 610\,MHz.
	See Figure~\ref{fig:f1929}.
    \label{fig:f1744}}
\end{figure}

PSR~B1744$-$24A, in the globular cluster Terzan 5, is a rare eclipsing
pulsar.  In Figure~\ref{fig:f1744}(b) we present the profile resulting
from 330 minutes of observations acquired on four separate days and at
a wide range of orbital phases; these averaged data yield a maximum
linear polarization of 9\%.  Our observations do not support the
result of Xilouris \etal (1998), who found a linear polarization
fraction of 60\%.  The data of Xilouris \etal (1998) were taken during
a single observing session (\cite{xil98}); although we saw no similar
results in our four days' observations, it is possible that the linear
polarization of 1744-24 is variable.

\subsection{PSR~B1821$-$24}\label{sec:1821}

The pulse profile is complex, with multiple components and strong
linear intensity in at least two components.  The position angle is
nearly flat across the individual components, though there are offsets
between components which are not consistent with orthogonal mode
changes.  We find a possible RVM fit for the position-angle swing,
with $\alpha = 40.7\pm1.7^\circ$, $\beta =40\pm10^\circ$.  Our
result is consistent with the estimate of $\alpha = 50^\circ$, $\beta
=40^\circ$ given in Backer and Sallmen (1997), \nocite{bs97} though the
phase of steepest swing is offset by roughly $180^\circ$.  However, when
both $\alpha$ and $\beta$ are near $45^\circ$, the RVM curve is nearly
symmetric, thus this discrepancy is not as important as it may seem.
We find no evidence in our 610\,MHz data for mode-changing as
described in Backer and Sallmen (1997); however, Backer and Sallmen
found no such mode-changes below 1395\,MHz.

\subsection{PSR~J1911$-$1114}\label{sec:1911}

There is some frequency evolution apparent in the presented profiles:
a trailing component becomes stronger at 610\,MHz than at 410\,MHz,
and there is a reduction in the intensities of both linear and
circular polarizations.  The strong circular polarization is
indicative of a core component, and the small trailer may be a conal
outrider.  Taken together, the polarization and the frequency
development suggest that this MSP may fit into the classification
scheme based on slower pulsars.  The position angle shows orthogonal
mode switching at both frequencies.
 
\subsection{PSR~B1937$+$21}\label{sec:1937}

This isolated object was the first millisecond pulsar discovered
(\cite{bkh+82}), and it is still the fastest known.  Its polarimetric
properties have previously been studied at multiple frequencies
(\cite{sti83,sc83,als83,sbc+84,ts90,sal98}).  Our results agree 
well with those published
earlier, and offer considerable information about the position angle
behavior at the leading edge of the main pulse at 1414\,MHz.  The
main-pulse peak linear polarization decreases from 75\% at 610\,MHz to
49\% at 1414\,MHz; the position angle is nearly flat across both the
main pulse and the interpulse, with evidence for orthogonal mode
changes at the leading edge of the main pulse at 1414\,MHz and of both
components at 610\,MHz.  Both components broaden as a result of
scattering at lower frequencies. As the main pulse and interpulse are
separated by nearly $180^\circ$ of phase, it is sensible to interpret the
emission as coming from an orthogonal rotator.  As mentioned in
Thorsett and Stinebring (1990), the observed pulse widths are much
narrower than those predicted by the empirical relation between the
spin period and magnetic inclination (\cite{ran90}).

\subsection{PSR~J2145$-$0750}\label{sec:2145}

The profile of this IMBP contains at least three components, with
bridged emission between the two strongest.  There is considerable
frequency evolution: the main pulse component strengthens relative to
the others at higher frequencies.  Along with the sense-reversing
circular polarization in the last component, this suggests that the
last component is core emission, while the ``main pulse'' is conal.
However, it is not clear how to interpret the precursor in this
picture.  The linear intensity is strongest at low frequencies, and
the accompanying position-angle swing is extremely complex, with
multiple orthogonal mode changes.  Even with the mode changes
subtracted, the RVM is not a good fit to the 410\,MHz or 610\,MHz
profiles.  At 1414\,MHz, we find that the linear intensity drops to
very low levels.  In contrast, Xilouris \etal (1998) present a profile
with reasonably strong linear polarization and a shallow
position-angle swing.  Sallmen (1998) reports a similar profile
(strong linear intensity, shallow position-angle swing) derived from a
single observation at 820\,MHz at Green Bank.  It seems therefore that
the linear emission is highly variable, further challenging models of
the pulsar emission mechanisms.

\section{Conclusion} \label{sec:concl}

Millisecond pulsars share a number of polarization characteristics
with slow pulsars; in particular, high degrees of linear and circular
polarization, well-defined position-angle swings, and orthogonal mode
switching appear to be common in both classes.  However, very few MSPs
have profiles which evolve following the predictions of the
slow-pulsar phenomenological model or have position-angle swings which
fit the rotating vector model.  Our results add to the evidence that
recycled pulsars have more complex polarimetric properties than
younger pulsars.

\acknowledgements

We thank David Nice, Jon Bell and Christopher Scaffidi for assistance
with observations, and Andrew Lyne, Joe Taylor and Michael Kramer for
valuable discussions.  I. H. S. received support from an NSERC 1967
fellowship. S. E. T. is an Alfred P. Sloan Research Fellow.  F. C. is
a Marie Curie Fellow.

\clearpage

\begin{figure}
    \plotone{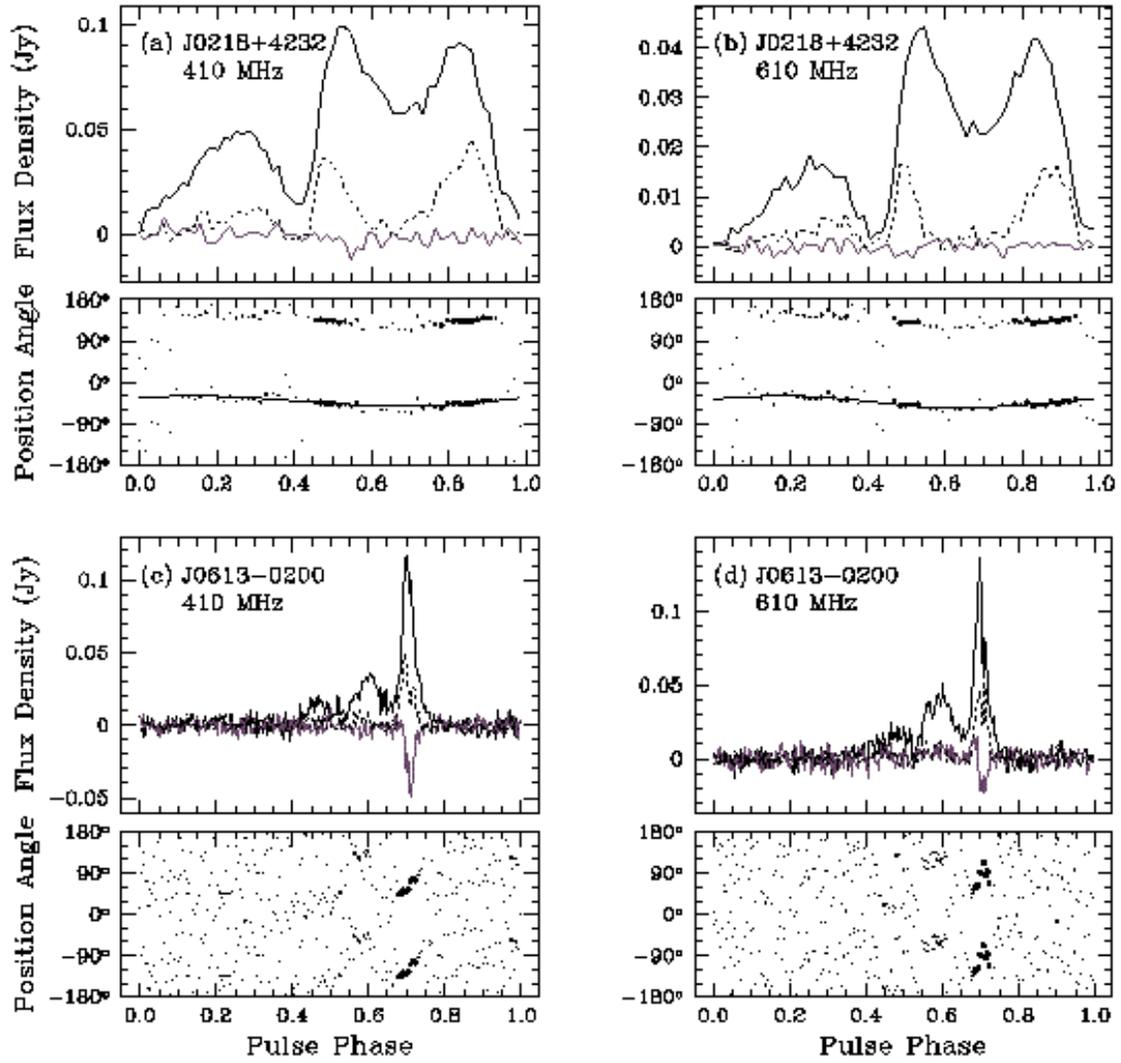}
    \figcaption
    {Pulse profiles for (a) PSR J0218+4232 at 410\,MHz,
	(b) PSR J0218+4232 at 610\,MHz,
	(c) PSR J0613$-$0200 at 410\,MHz,
	(d) PSR J0613$-$0200 at 610\,MHz.
	See Figure~\ref{fig:f1929}.
    \label{fig:f0218}}
\end{figure}

\begin{figure}
    \plotone{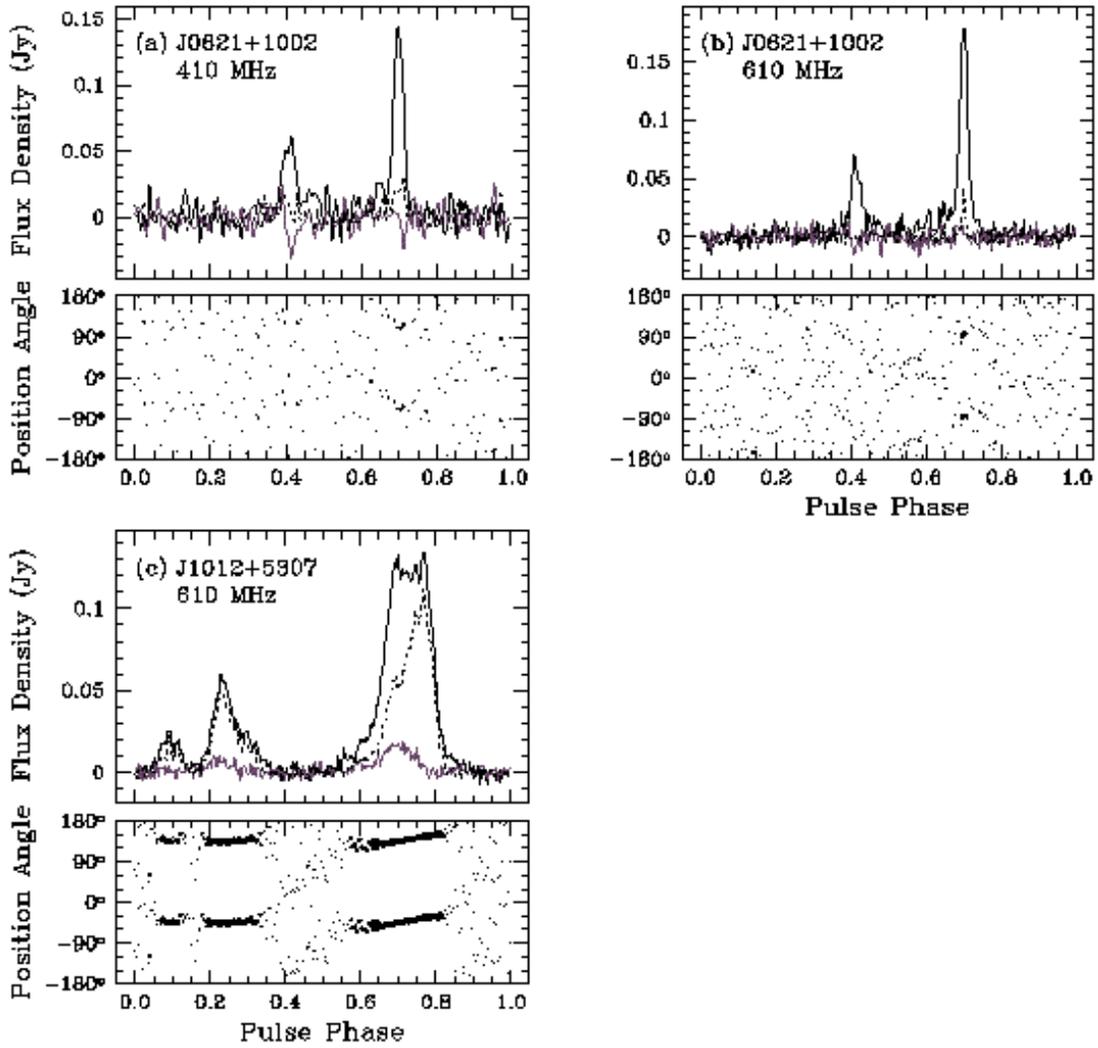}
    \figcaption
    {Pulse profiles for (a) PSR J0621+1002 at 410\,MHz,
	(b) PSR J0621+1002 at 610\,MHz,
	(c) PSR J1012+5307 at 610\,MHz.
	See Figure~\ref{fig:f1929}.
    \label{fig:f0621}}
\end{figure}
 
\begin{figure}
    \plotone{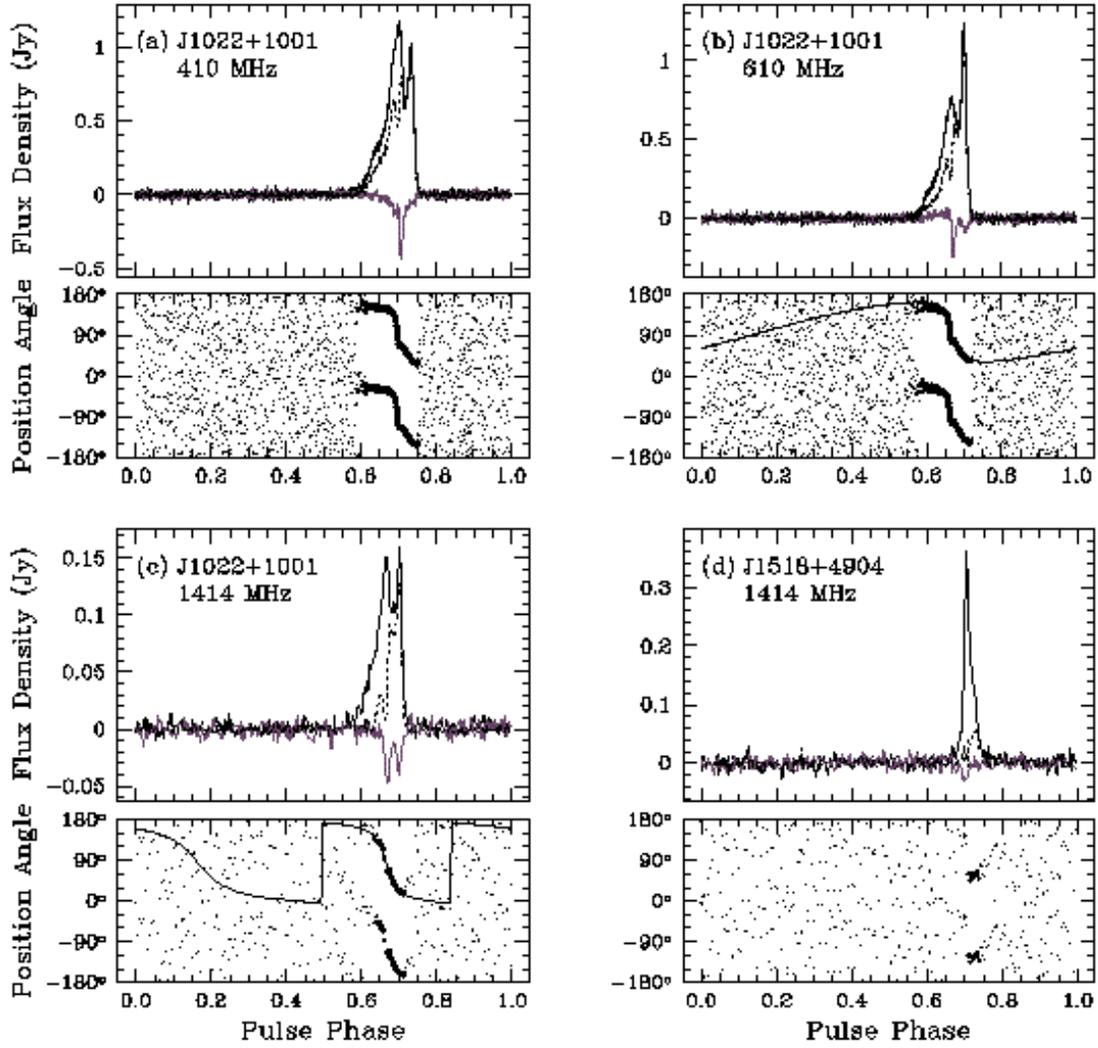}
    \figcaption
    {Pulse profiles for (a) PSR J1022+1001 at 410\,MHz,
	(b) PSR J1022+1001 at 610\,MHz,
	(c) PSR J1022+1001 at 1414\,MHz,
	(d) PSR J1518+4904 at 610\,MHz.
	See Figure~\ref{fig:f1929}.
    \label{fig:f1022}}
\end{figure}

\begin{figure}
    \plotone{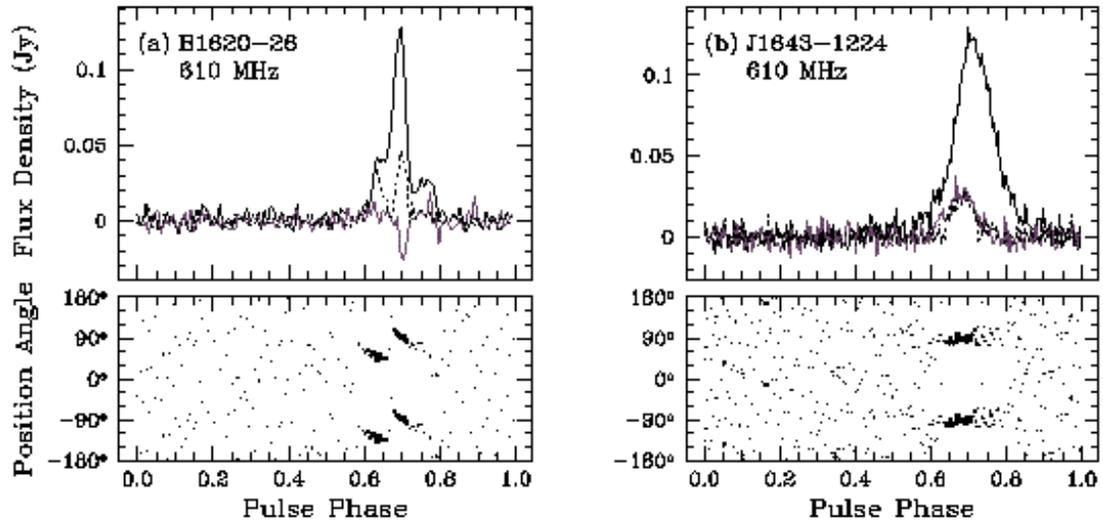}
    \figcaption
    {Pulse profiles for 
	(a) PSR B1620$-$26 at 610\,MHz,
	(b) PSR J1643$-$1224 at 610\,MHz.
	See Figure~\ref{fig:f1929}.
    \label{fig:f1620}}
\end{figure}

\begin{figure}
    \plotone{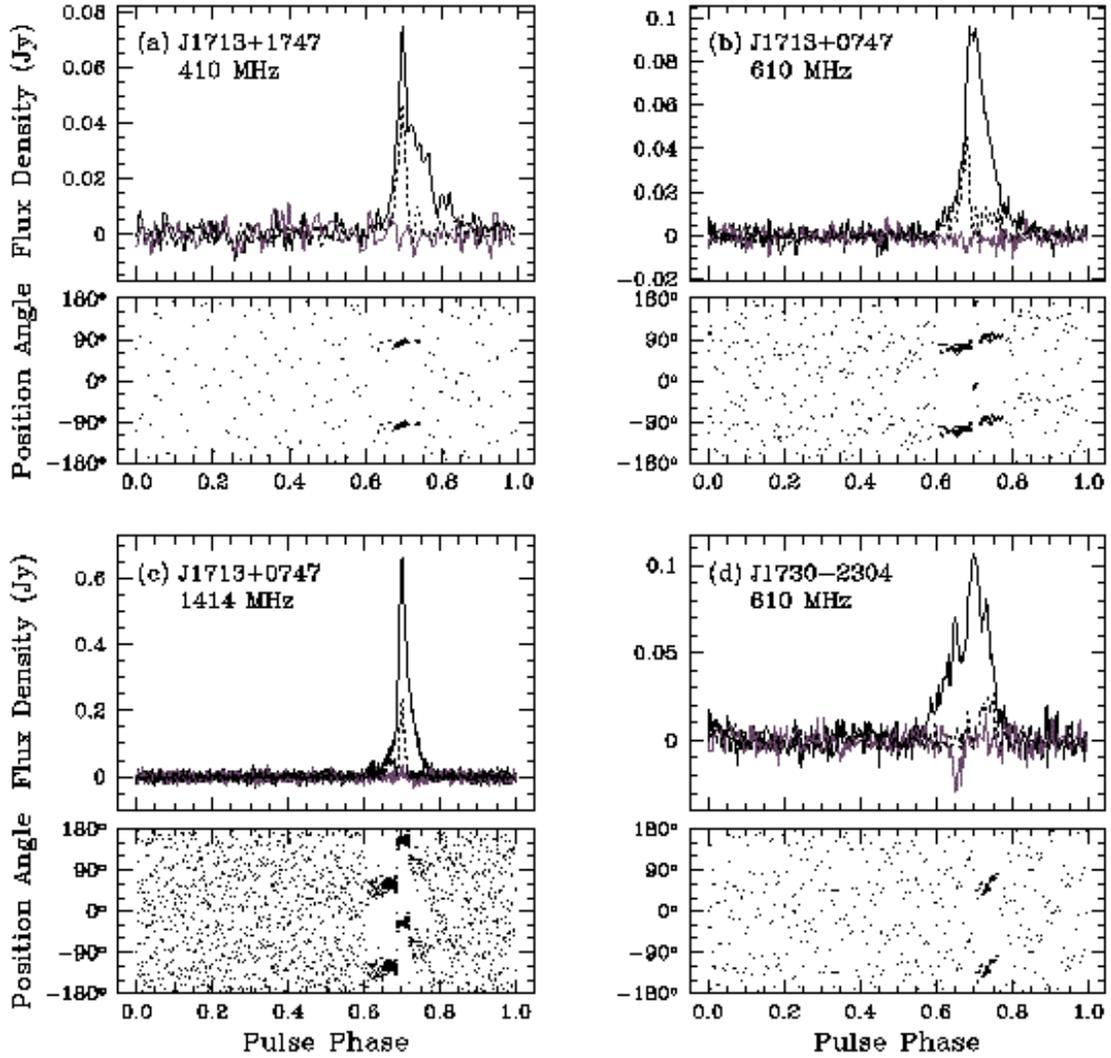}
    \figcaption
    {Pulse profiles for (a) PSR J1713+0747 at 410\,MHz,
	(b) PSR J1713+0747 at 610\,MHz,
	(c) PSR J1713+0747 at 1414\,MHz,
	(d) PSR J1730$-$2304 at 610\,MHz.
	See Figure~\ref{fig:f1929}.
    \label{fig:f1713}}
\end{figure}
 
\begin{figure}
    \plotone{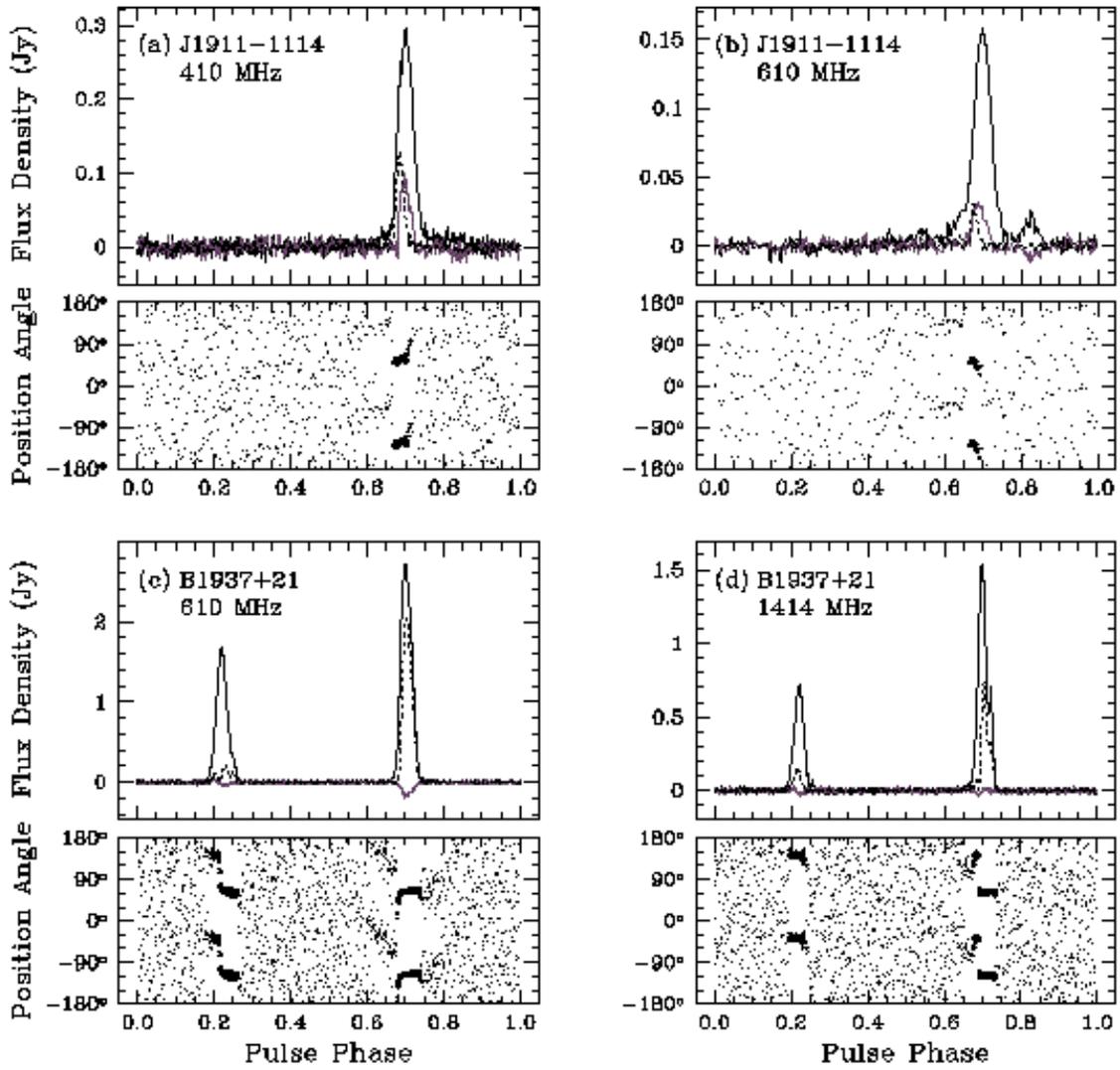}
    \figcaption
    {Pulse profiles for (a) PSR J1911$-$1114 at 410\,MHz,
	(b) PSR J1911$-$1114 at 610\,MHz,
	(c) PSR B1937+21 at 610\,MHz,
	(d) PSR~B1937+21 at 1414\,MHz.
	See Figure~\ref{fig:f1929}.
    \label{fig:f1911}}
\end{figure}

\begin{figure}
    \plotone{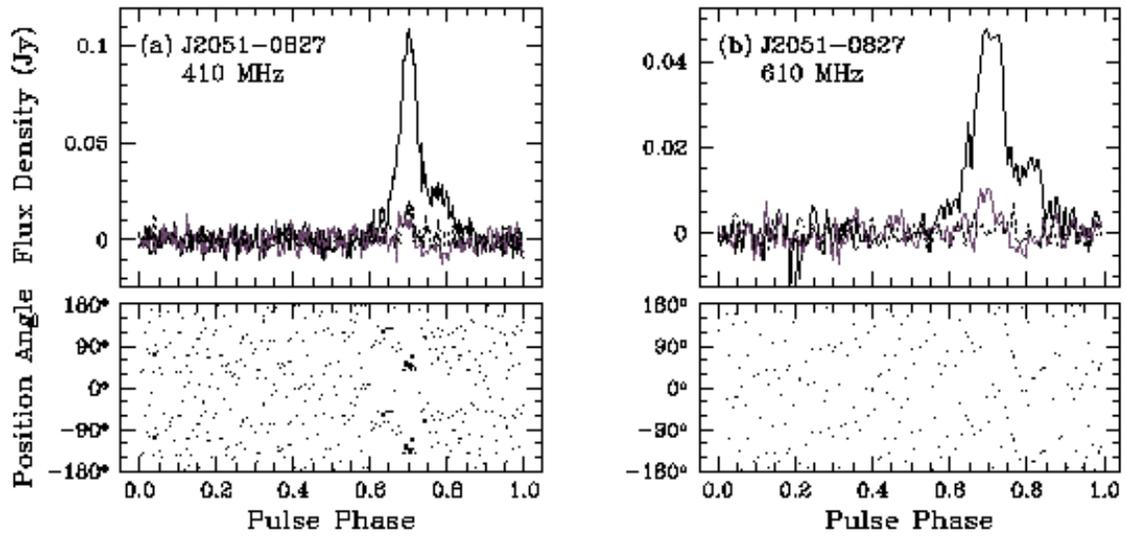}
    \figcaption
    {Pulse profiles for (a) PSR J2051$-$0827 at 410\,MHz,
	(b) PSR J2051$-$0827 at 610\,MHz.	
	See Figure~\ref{fig:f1929}.
    \label{fig:f2051}}
\end{figure}

\begin{figure}
    \plotone{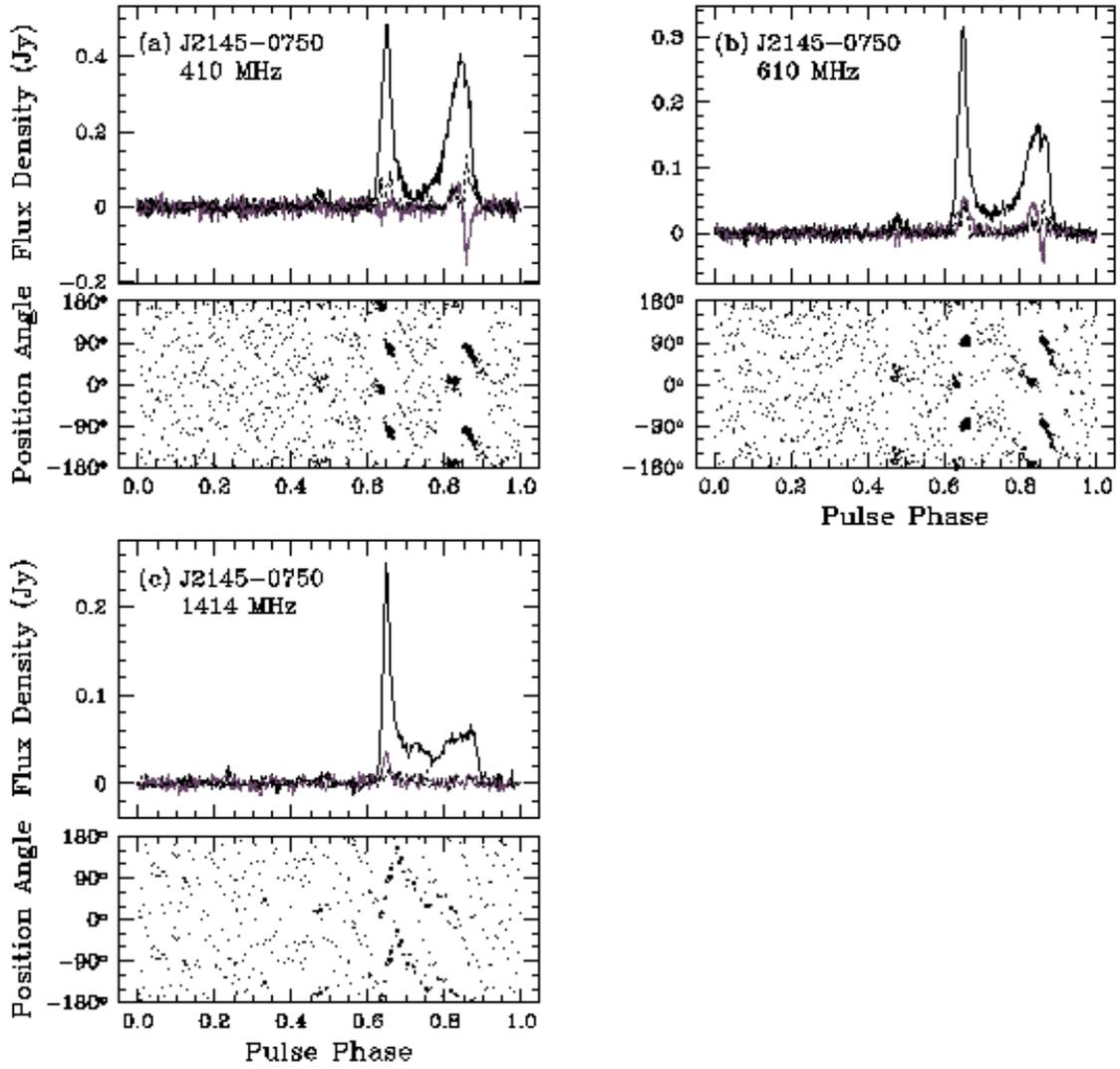}
    \figcaption
    {Pulse profiles for (a) PSR J2145$-$0750 at 410\,MHz,
	(b) PSR J2145$-$0750 at 610\,MHz,
	(c) PSR J2145$-$0750 at 1414\,MHz.
	See Figure~\ref{fig:f1929}.
    \label{fig:f2145}}

\end{figure}
 
\clearpage

\begin{table*}[t]
\caption {Rotating vector model fits for PSR~B1929$+$10\tablenotemark{a}.\label{tab:1929}}
\vspace{2mm}
 
\begin{center}
\begin{tabular}{r@{\hspace{0.3in}}r@{$\pm$}lr@{$\pm$}lr@{$\pm$}lc}
\tableline
\tableline
\multicolumn{1}{c}{Frequency}	&
\multicolumn{2}{c}{}	&
\multicolumn{2}{c}{} &
\multicolumn{2}{c}{Phase of Steepest} &
\multicolumn{1}{c}{Reduced} \\
\multicolumn{1}{c}{(MHz)} &
\multicolumn{2}{c}{$\alpha$ ($^\circ$)} &
\multicolumn{2}{c}{$\beta$ ($^\circ$)} &
\multicolumn{2}{c}{PA Slope ($^\circ$)\tablenotemark{b}} &
\multicolumn{1}{c}{$\chi^2$} \\
\tableline
410 & \hspace{0.01in} $41$&$8$ & $21$&$4$ & \hspace{0.25in}$-12.5$&$1.0$ & 1.03 \\
610 & $51$&$3$ & $35$&$3$ &$-11.2$&$1.4$ & 1.00 \\
 1414 & $61$&$2$ & $39$&$2$ & $-5$&$2$ & 1.02 \\
\tableline
\end{tabular}
\end{center}

\begin{raggedright}
\tablenotetext{a}{Only those position angle points for which the 
corresponding linear intensity is 3 or more times the off-pulse
noise are included.}
\tablenotetext{b}{The phase of steepest position-angle slope is
given relative to the peak of the profile.}
\end{raggedright}
\end{table*}

\begin{deluxetable}{lrlrrc}
\tablecolumns{6}
\tablecaption{Pulsars observed at Jodrell Bank.\label{tab:ppdot}}
\tablehead{
\multicolumn{1}{c}{Pulsar } &
\multicolumn{1}{c}{$P$}	&
\multicolumn{1}{c}{$\dot P$} &
\multicolumn{1}{c}{$\log B$} &
\multicolumn{1}{c}{$\log \tau_{\rm c}$}  &
\multicolumn{1}{c}{Reference\tablenotemark{a}} \nl
 &
\multicolumn{1}{c}{(ms)} &
\multicolumn{1}{c}{($10^{-18}$)} &
\multicolumn{1}{c}{(G)} &
\multicolumn{1}{c}{(yr)} & 
}
\startdata
J0034$-$0534\dotfill 	& 1.88	& 0.0051  	& 8.00	& 9.77	&  \nl
J0218$+$4232\dotfill 	& 2.32	& 0.08 		& 8.64	& 8.66	&  \nl
J0613$-$0200\dotfill 	& 3.06	& 0.0096 	& 8.24	& 9.70	&  \nl
J0621$+$1002\dotfill 	& 28.85	& 0.04		& 9.04	& 10.06	& 1,2\nl
J1012$+$5307\dotfill 	& 5.26	& 0.015		& 8.45	& 9.74	&  \nl
J1022$+$1001\dotfill 	& 16.45	& 0.04		& 8.91	& 9.81	& 1 \nl
J1518$+$4904\dotfill 	& 40.93	& 0.027		& 9.03	& 10.38	& 3 \nl
B1620$-$26\dotfill	& 11.08	& 0.82		& 9.48	& 8.33	&  \nl
J1643$-$1224\dotfill	& 4.62	& 0.018		& 8.47	& 9.61	&  \nl
J1713$+$0747\dotfill	& 4.57	& 0.0085 	& 8.30	& 9.93	&  \nl
J1730$-$2304\dotfill 	& 8.12	& 0.020		& 8.61	& 9.81	&  \nl
J1744$-$1134\dotfill 	& 4.07	& 0.0086 	& 8.28	& 9.87	& 4 \nl
B1744$-$24A\dotfill	& 11.56	& $-$0.019 	&  	&	&  \nl
B1820$-$30A\dotfill	& 5.44	& 3.38		& 9.64	& 7.41	&  \nl
B1821$-$24\dotfill	& 3.05	& 1.61		& 9.35	& 7.48	&  \nl
J1911$-$1114\dotfill 	& 3.63	& 0.013		& 8.34	& 9.65	& 5 \nl
B1937$+$21\dotfill	& 1.56	& 0.11		& 8.62	& 8.35	&  \nl
J2051$-$0827\dotfill 	& 4.51	& 0.013		& 8.39	& 9.74	& 6 \nl
J2145$-$0750\dotfill 	& 16.05	& 0.030		& 8.85	& 9.93	& 2 \nl
\enddata
\tablenotetext{a}{All $P$ and $\dot P$ values from Taylor \etal (1995),
\nocite{tmlc95} except:
1) Camilo \etal (1996), \nocite{cnst96} 
2) Camilo (1997), \nocite{cam97} 
3) Sayer \etal (1997), \nocite{snt97}
4) Bailes \etal (1997), \nocite{bjb+97}
5) Lorimer \etal (1996), \nocite{llb+96} 
6) Stappers \etal (1996). \nocite{sbl+96}}
\end{deluxetable}

\clearpage

\begin{deluxetable}{lrrrrc}
\tablecaption{Parameters of polarimetry observations.\label{tab:res}}
\tablecolumns{6}
\tablehead{
\multicolumn{1}{c}{Pulsar } \hspace{0.7in}& 
\multicolumn{1}{c}{Frequency}	&
\multicolumn{1}{c}{Resolution}	&
\multicolumn{1}{c}{Integration time} &
\multicolumn{1}{c}{Integration time} &
\multicolumn{1}{c}{Previous} \nl
 & 
\multicolumn{1}{c}{(MHz)} &
\multicolumn{1}{c}{($\mu$s)} &
\multicolumn{1}{c}{(profile) (min)} &
\multicolumn{1}{c}{(flux) (min)} &
\multicolumn{1}{c}{Polarimetry\tablenotemark{a}} \nl
}
\startdata
J0034$-$0534\dotfill 	& 410 & 15  	& 180 	& 270 &      \nl
J0218$+$4232\dotfill 	& 410 & 36  	& 60 	& 60 &       \nl
			& 610 & 36  	& 180 	& 300 &         \nl
J0613$-$0200\dotfill 	& 410 & 12  	& 180 	& 180 &       \nl
			& 610 & 12  	& 90 	& 150 &  1     \nl 
J0621$+$1002\dotfill 	& 410 & 225  	& 30 	& 30  &      \nl
			& 610 & 113 	& 60 	& 90  &       \nl 
J1012$+$5307\dotfill 	& 610 & 21  	& 300 	& 360  & 1    \nl
J1022$+$1001\dotfill 	& 410 & 16  	& 30 	& 30  &  2     \nl
			& 610 & 16  	& 30 	& 180 &  1     \nl
			& 1414 & 64   	& 50 	& 185 &   1,3     \nl
J1518$+$4904\dotfill 	& 610 &  160 	& 20 	& 20  &         \nl
B1620$-$26\dotfill	& 610 & 87  	& 60 	& 120  &   1     \nl 
J1643$-$1224\dotfill	& 610 & 18  	& 90 	& 90 &       \nl
J1713$+$0747\dotfill	& 410 &  36 	& 240 	& 240  &        \nl
			& 610 &  18 	& 300 	& 390  &  1     \nl
			& 1414 &  4 	& 30 	& 140  &  1,3     \nl 
J1730$-$2304\dotfill 	& 610 &  32 	& 120 	& 120  &  1       \nl
J1744$-$1134\dotfill 	& 610 &  16 	& 30 	& 30   &      \nl
B1744$-$24A\dotfill	& 1414 &  181 	& 330 	& 420 &   3     \nl 
B1820$-$30A\dotfill	& 610 & 42  	& 90 	& 90  &     \nl
B1821$-$24\dotfill	& 610 & 12  	& 90 	& 90   &   1,4    \nl
J1911$-$1114\dotfill 	& 410 & 7  	& 270 	& 270  &        \nl
			& 610 & 14  	& 210 	& 330 &        \nl
B1937$+$21\dotfill	& 610 & 1.5  	& 90 	& 250 &  1       \nl
			& 1414 & 1.5  	& 105 	& 620 &  1,5     \nl
J2051$-$0827\dotfill 	& 410 &  18 	& 110 	& 140 &        \nl
			& 610 &  35 	& 120 	& 300  &         \nl
J2145$-$0750\dotfill 	& 410 &  31 	& 60 	& 60   &       \nl
			& 610 &  31 	& 120 	& 120  &  1      \nl
			& 1414 & 63  	& 30 	& 130  &  1,3     \nl
\enddata
\tablenotetext{a}{References: 1) Sallmen (1998), \nocite{sal98} 
2) Shrauner (1997), \nocite{shr97} 3) Xilouris \etal
(1998), \nocite{xkj+98} 4) Backer and Sallmen (1997), \nocite{bs97}
5) Thorsett and Stinebring (1990). \nocite{ts90}}
\end{deluxetable}

\clearpage

\begin{deluxetable}{lrrrrrrr}
\tablecaption{Pulse widths, mean polarization intensities and mean
fluxes\tablenotemark{a}.\label{tab:pol}}
\tablehead{
\multicolumn{1}{c}{Pulsar } & 
\multicolumn{1}{c}{Frequency}	&
\multicolumn{1}{c}{$w_{50}$}	&
\multicolumn{1}{c}{$w_{10}$} &
\multicolumn{1}{c}{$w_{\rm e}$} &
\multicolumn{1}{c}{Mean} &
\multicolumn{1}{c}{Mean} &
\multicolumn{1}{c}{Mean} \nl
 & 
\multicolumn{1}{c}{(MHz)} &
\multicolumn{1}{c}{(mP)} &
\multicolumn{1}{c}{(mP)} &
\multicolumn{1}{c}{(mP)} &
\multicolumn{1}{c}{L (\%)} &
\multicolumn{1}{c}{$|$V$|$ (\%)} &
\multicolumn{1}{c}{Flux (mJy)} \nl
}
\startdata
J0034$-$0534\dotfill 	& 410 & 344 	& 524 	& 266 	& 0 & 18 & 24 \nl
J0218$+$4232\dotfill 	& 410 & 452 	& 	& 477	& 24 & 5 & 47 \nl
			& 610 & 435 	& 	& 423	& 22 & 4 & 22 \nl
J0613$-$0200\dotfill 	& 410 & 34 	& 327	& 75	& 20 & 28 & 9.2 \nl
			& 610 & 30 	& 315	& 72	& 22 & 24 & 6.9 \nl
J0621$+$1002\dotfill 	& 410 & 31 	& 	& 63	& 13 & 39 & 9.5 \nl
			& 610 & 24	& 352	& 55	& 6 & 22 & 9.3 \nl
J1012$+$5307\dotfill 	& 610 & 137 	& 	& 193	& 61 & 12 & 24 \nl
J1022$+$1001\dotfill 	& 410 & 69 	& 137	& 72	& 70 & 13 & 75 \nl
			& 610 & 56 	& 124	& 51	& 71 & 10 & 22 \nl
			& 1414 & 66  	& 132	& 73	& 40 & 25 & 5.3 \nl 
J1518$+$4904\dotfill 	& 610 &  17	& 55	& 27	& 17 & 17 & 8.7 \nl
B1620$-$26\dotfill	& 610 & 39 	& 176	& 72	& 30 & 26 & 9.0 \nl
J1643$-$1224\dotfill	& 610 & 117 	& 436	& 129	& 12 & 22 & 16 \nl
J1713$+$0747\dotfill	& 410 &  42	& 	& 71	& 27 & 29 & 6.8 \nl
			& 610 &  61	& 210	& 78	& 26 & 12 & 6.8 \nl
			& 1414 &  24	& 92	& 37	& 22 & 7 & 7.9 \nl
J1730$-$2304\dotfill 	& 610 &  97	& 322	& 97	& 11 & 21 & 11 \nl
J1744$-$1134\dotfill 	& 610 &  33	& 82	& 46	& 64 & 19 & 13 \nl
B1744$-$24A\dotfill	& 1414 &  59	& 112	& 63	& 7 & 14 & 4.6 \nl
B1820$-$30A\dotfill	& 610 & 63 	& 180	& 78	& 0 & 18 & 7.1 \nl
B1821$-$24\dotfill	& 610 & 42 	& 	& 113	& 41 &	26 & 17 \nl 
J1911$-$1114\dotfill 	& 410 & 45 	& 91	& 51	& 20 & 24 & 15 \nl
			& 610 & 51 	& 125	& 68 	& 7 & 19 & 8.1 \nl
B1937$+$21\dotfill	& 610 & 35 	& 56	& 57	& 42 & 5 & 131 \nl
			& 1414 & 24 	& 55	& 44	& 29 & 3 & 24 \nl
J2051$-$0827\dotfill 	& 410 &  48	& 252	& 75	& 8 & 15 & 7.9 \nl 
			& 610 &  98	& 296	& 127	& 1 & 18 & 4.2 \nl
J2145$-$0750\dotfill 	& 410 &  234	& 265	& 96	& 13 & 14 & 46 \nl
			& 610 &  215	& 261	& 80	& 11 & 17 & 19 \nl
			& 1414 & 21 	& 261	& 67	& 5 & 18 & 6.6 \nl
\enddata
\tablenotetext{a}{Uncertainties discussed in text.}
\end{deluxetable}
 \clearpage

\begin{table*}[t]
\caption 
{Rotating vector model fits\tablenotemark{a}\label{tab:rvm}}
\vspace{2mm}
 
\begin{center}
{
\renewcommand{\tabcolsep}{1.4mm}
\begin{tabular}{lr@{\hspace{0.3in}}r@{}c@{}lr@{}c@{}lr@{}c@{}lc}
\tableline
\tableline
\multicolumn{1}{c}{Pulsar } \hspace{0.3in}& 
\multicolumn{1}{c}{Frequency}	&
\multicolumn{3}{c}{}	&
\multicolumn{3}{c}{} &
\multicolumn{3}{c}{Phase of Steepest} &
\multicolumn{1}{c}{Reduced} \\
 & 
\multicolumn{1}{c}{(MHz)} &
\multicolumn{3}{c}{$\alpha$ ($^\circ$)} &
\multicolumn{3}{c}{$\beta$ ($^\circ$)} &
\multicolumn{3}{c}{PA Slope ($^\circ$)\tablenotemark{b}} &
\multicolumn{1}{c}{$\chi^2$} \\
\tableline
J0218$+$4232\dotfill 	& 410 & $8$&$\pm$&$11$ &
\multicolumn{3}{c}{undetermined} & \hspace{0.45in} $-39$&$\pm$&$5$ & 0.87 \\\vspace{0.2in}
			& 610 & $8$&$\pm$&$15$ & \multicolumn{3}{c}{undetermined} & $-45$&$\pm$&$5$ & 1.03 \\
J1022$+$1001\dotfill 	& 610 & $140$&$\pm$&$16$ & $-4.9$&$\pm$&$1.8$ &
$-12.2$&$\pm$&$0.2$ &  1.06 \\\vspace{0.2in}
			& 1414 & $83$&$\pm$&$27$ & $-7.1$&$\pm$&$0.4$ &
$-13.3$&$\pm$&$0.5$ &  1.04 \\
B1821$-$24\dotfill	& 610 & $40.7$&$\pm$&$1.7$ & $40$&$\pm$&$10$ &
$-167.1$&$\pm$&$1.9$  & 1.04 \\
\tableline
\end{tabular}
}
\end{center}
\begin{raggedright}
\tablenotetext{a}{Only those position angle points for which the 
corresponding linear intensity is 3 or more times the off-pulse
noise are included.}
\tablenotetext{b}{The phase of steepest position-angle slope is given
relative to the peak of the profile.}
\end{raggedright}
\end{table*}

\end{document}